\documentclass[pre,preprint,superscriptaddress,amsmath,amssymb,floatfix]
{revtex4}
\usepackage{graphics}
\usepackage[dvips]{graphicx}
\begin{document}
%%%%%%%%%%%%%%%%%%%%%%%%%%%%%%%%%%%%%%%%%%%
\title{Correlation functions in ionic liquid
 at coexistence with ionic crystal. Results of the Brazovskii-type
 field theory\footnote{Dedicated to Prof. R. Evans on
the occasion of his 60th birthday}}
 \author{O. Patsahan}
\affiliation{Institute for Condensed Matter Physics of the National
Academy of Sciences of Ukraine, 79011 Lviv, Ukraine}
 \author{ A. Ciach}
\address{Institute of Physical Chemistry,
 Polish Academy of Sciences, 01-224 Warszawa, Poland}
 \date{\today} 
%%%%%%%%%%%%%%%%%%%%%%%%%%%%%%%%%%%%%%%%%%%%%%%%%%%%%%%%%%%%%%
\begin{abstract}
Correlation functions in the restricted primitive model are calculated
within a field-theoretic approach in the one-loop self-consistent
Hartree approximation. The correlation functions exhibit damped
oscillatory behavior as found before in the Gaussian approximation
[Ciach at. al., J. Chem. Phys. {\bf 118}, 3702 (2003)]. The
fluctuation contribution leads to a renormalization of both the
amplitude and the decay length of the correlation functions. The
renormalized quantities show qualitatively different behavior than
their mean-field (MF) counterparts. While the amplitude and the decay
length both diverge in MF when the $\lambda$-line is approached, the
renormalized quantities remain of order of unity in the same dimensionless
units down to the coexistence with the ionic crystal. Along the line
of the phase transition the decay length and the period of
oscillations are independent of density, and their values in units of
the diameter of the ions are $\alpha_0^{-1}\approx 1$ and
$2\pi/\alpha_1\approx 2.8$ respectively.
\end{abstract} 
\maketitle
%%%%%%%%%%%%%%%%
 \section{Introduction}

Distribution of charges in ionic systems has been a subject of 
considerable interest for many years. In theoretical studies of
systems such as molten salts, electrolytes or ionic liquids the
interaction potential is often approximated by the restricted
primitive model (RPM), where identical hard spheres carry positive or
negative charges with equal magnitude.  Earlier studies concentrated on
correlation functions for charges in ion-dilute systems
\cite{debye:23:0,fisher:94:0,stell:95:0}, and later the ion-dense systems
were also
investigated~\cite{stillinger:68:0,leote:94:0,fisher:94:0,stell:95:0,evans:95:0,evans:94:0,evans:96:0,leote:99:0}.
However, the question of the form of the correlation
functions in the liquid (fused salt) at the coexistence with
the ionic crystal remain open. Whereas the correlation functions in
a liquid can be quite accurately described within liquid theories such
as the mean-spherical approximation
(MSA)~\cite{leote:94:0,stell:95:0}, the transition to a crystalline
phase is not predicted by the liquid theories. The liquid-solid
transition in the RPM was determined within the density functional
theory of freezing in Ref.~\cite{barrat:87:0}, but the charge-charge
correlation function is not discussed in this work. Both the phase
transition and the correlation functions can be found on the same
level of approximation within the framework of the field theory for
the RPM introduced in Ref.\cite{ciach:00:0}.  In this work we
calculate the correlation functions  in the RPM by
applying the Brazowskii-type approximation. We pay particular
attention to the decay length, the period of damped oscillations and
the amplitude along the phase coexistence with an ionic crystal that
was recently found in Ref.~\cite{ciach:06:2}.
%%%%%%%%%%%%%%%%%%%%%%%%%%%%%%%%

The behavior of the charge-charge correlation function $h_D(x)$ depends 
on thermodynamic conditions. In Fig.1 a portion of the 
density-temperature 
phase diagram, as obtained in different 
theoretical~\cite{stell:95:0,fisher:94:0} and
simulation~\cite{vega:03:0} studies, is shown
 schematically. The solid lines represent
 boundaries of stability of the uniform fluid phase, and regions 
corresponding to different 
qualitative properties of
$h_D(x)$ are separated by the dashed lines. 

The Kirkwood line (K) separates the high-temperature/low-density
region corresponding to an asymptotic monotonic decay of correlations,
from the low-temperature/high-density region corresponding to
exponentially damped oscillatory behavior of $h_D(x)$. In theories
of the mean-field (MF) type the Kirkwood
line is found to be a straight line $S=S_K=const$, where
\begin{equation}
\label{SSMF}
S\equiv \frac{T^*}{\rho^*_0}.
\end{equation}
In the above $\rho^*_0$ is the dimensionless  number-density (volume measured in $\sigma^3$ units, where $\sigma$ is the ion-diameter), and
\begin{equation}
\label{T*}
 T^*=\frac{D\sigma kT}{e^2}
\end{equation}
 is the temperature
in standard reduced units ($D$ is the
dielectric constant of the solvent and $e$ is the charge). In different theories  the value of $S_K$ is somewhat
different, and typically
$S_K\sim 10$. There is no controversy concerning this line, but there are
no general reasons for this line to be straight beyond
MF~\cite{stell:95:0}.
%%%%%%%%%%%%%%%%%%%%%
\begin{figure}
\includegraphics[scale=0.5]{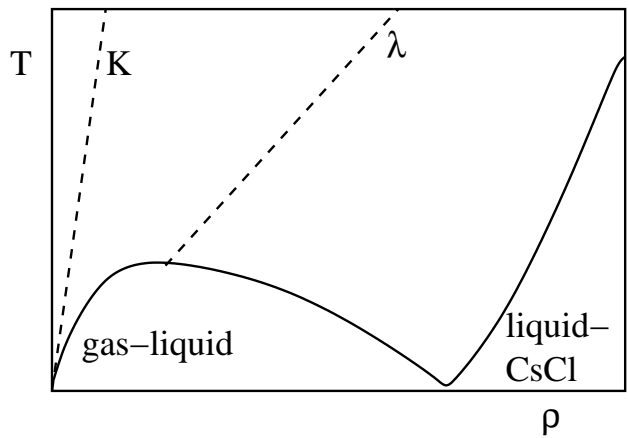}
\caption{A portion of the density-temperature phase diagram for the RPM as obtained in MF, shown
 schematically. Solid lies separate the uniform fluid from the 
two-phase regions. Dashed lines denoted by 'K' and '$\lambda$' 
represent the Kirkwood line and the $\lambda$-line respectively.
 $T$ and $\rho$ are in arbitrary units.}
\end{figure}

%%%%%%%%%%%%%%%%%%%%%%%%%%%%%%%%%%%%%
 When the $\lambda$-line is approached from the high-temperature side,
 the decay vent of correlations
 diverges~\cite{ciach:00:0,patsahan:04:0,outhwaite:75:0}. The amplitude of
 the correlation function diverges~\cite{ciach:03:1} as well on the MF
 level of the field theory~\cite{ciach:00:0}. The low-temperature side
 of this line corresponds to a charge-ordered structure, where the
 nearest-neighbors (nn) are oppositely charged, and the $\lambda$-line
 represents a continuous transition between the uniform and the
 charge-ordered phases. The slope of this line shows a strong
 dependence on approximations and assumptions made in different
 theories~\cite{outhwaite:75:0}. In particular, in the MF approximation for
 the field theory the $\lambda$-line is $S=S_{\lambda}\approx
 1.6$~\cite{ciach:00:0}, whereas the MSA yields
 $S_{\lambda}=0$~\cite{leote:94:0}. The charge-ordered liquid phase
 was observed neither experimentally nor in simulations. A natural
 conclusion would be to question validity of theories that predict
 the existence of the $\lambda$-line. However, the $\lambda$-line of
 continuous transitions to the charge-ordered phase was observed in
 simulations ~\cite{panag:99:0,diehl:03:0,dickman:99:0} of the lattice
 version of the RPM (LRPM), in agreement with predictions of the
 lattice version of the field theory~\cite{ciach:01:1}. Some other
 theories~\cite{kobelev:02:0,brognara:02:0,artyomov:03:0} yield for the LRPM similar
 results. The MSA, however, predicts an absence of such a transition
 in the LRPM. One might argue that the MSA theories are correct only
 in the continuum space, and the field theory yields correct results
 only on the lattice. This is indeed the case for the field theory in
 the MF approximation. Beyond MF, with the charge fluctuations
 accounted for in a Brazovskii-type
 approximation~\cite{ciach:03:0,ciach:05:0,ciach:06:2}, the decay
 length of the charge-charge correlation function is found to be finite in the
 continuum-space RPM~\cite{ciach:06:2}. Thus, when the fluctuations
 are included in the field theory, there is no line of continuous transitions,
  in agreement with the MSA. At the same time the first-order transition to the
 ionic crystal is found~\cite{ciach:06:2} for the range of densities
 and temperatures that agree with simulations~\cite{vega:03:0}. Thus, within the field theory we can calculate the correlation functions at the coexistence with the ionic crystal.

The field-theory results~\cite{ciach:03:0,ciach:05:0,ciach:06:2} also
shed light on the nature of the $\lambda$-line that appears in MF.
Beyond the Kirkwood line there is no qualitative change of the {\it
average} structure of the liquid phase, which is represented by the
correlation function. On the other hand, the line $S\approx 1.6$
separates the phase diagram $(\rho^*_0,T^*)$ into two regions that
correspond to different forms of the {\it most probable} instantaneous
distributions of ions. At the high-temperature side of the line
$S=S_{\lambda}$ the distribution of ions is most probably random. For
$S<S_{\lambda}$, however, the most probable instantaneous
distributions of the ions correspond to charge-density waves with a
wavelength $\sim 2.5\sigma$, where $\sigma$ is the hard-sphere
diameter. According to the above prediction, in real-space
representation charge-ordered clusters with a nn distance $\sim
1.27\sigma$ should dominate over randomly distributed ions for
$S<S_{\lambda}$ in instantaneous charge distributions (i.e. in
microscopic states). Such structures were indeed seen in simulation
snapshots~\cite{panag:02:0,weis:98:0}. In MF the average density
distributions are approximated by the most probable
distributions. When in a particular system the average- and the
most-probable distributions agree on a qualitative level, the MF
results are qualitatively correct. This is not the case in the RPM,
therefore the line $S=S_{\lambda}$ is erroneously interpreted as a
line of continuous phase transitions in MF theories. On the other
hand, the occurrence of the $\lambda$-line on the MF level is an
indication of the first-order transition to the ionic crystal, because
beyond MF the first-order transition is found in the phase-space
region where the charge-ordered individual microstates are more
probable than the microstates with randomly distributed ions.

In this work we focus mainly on the charge-charge correlations for
$S<S_{\lambda}$, and calculate the number-number correlation function
as well. In Ref.\cite{ciach:03:1} the correlation functions are
calculated within the field-theoretic description developed in
Ref.\cite{ciach:00:0} only in the Gaussian approximation. As already
mentioned, at the $\lambda$-line the decay length and the
amplitude of the charge-charge correlation function both diverge. Our
purpose is to verify if the instantaneous order has some effect on the
correlation functions beyond the Gaussian approximation. The results
of Ref.~\cite{ciach:03:1} are extended beyond the Gaussian
approximation within the Brazovskii theory~\cite{brazovskii:75:0}.  It
turns out that at the coexistence with the ionic crystal, obtained in
Ref.\cite{ciach:06:2}, the amplitude and the decay length are quite
small. Our results show that large decay length and amplitude of the
charge-charge correlation function can be found only in the metastable
liquid.
%\end{document}
%%%%%%%%%%%%%%%%%%%%%%%%%%%%%%%%%%%%%%%%%%%
\section{Formalism}
\subsection{Field-theoretic description of the RPM}
 In the field theory derived for the RPM in
 Ref.\cite{ciach:00:0,ciach:05:0,ciach:04:1} the fluctuating fields
 are the local deviations from the most probable number- and charge
 densities, $\eta({\bf x})=\rho^*({\bf x})-\rho^*_0 =\rho^*_+({\bf
 x})+\rho^*_-({\bf x})-\rho^*_0$ and $\phi({\bf x})=\rho^*_+({\bf
 x})-\rho^*_-({\bf x})$ respectively. The most probable number density
 of ions is denoted by $\rho^*_0$, and the subscripts $+$ and $-$
 refer to cations and anions respectively.  Asterisks indicate that
 all densities are dimensionless (the unit volume is
 $\sigma^3$). $\phi$ is the charge density in $e/\sigma^3$ units.  We
 focus on systems that are globally charge-neutral,
\begin{equation}
\int_{\bf x}\phi({\bf x})=0,
\end{equation}
where in this paper we use the notation $\int_{\bf x}\equiv\int d{\bf
x}$. The fields $\eta({\bf x})$ and $\phi({\bf x})$ are thermally
excited with the probability density
\cite{ciach:00:0,ciach:06:0,ciach:05:0}
\begin{equation}
p[\phi,\eta]=\Xi^{-1}\exp\big(-\beta \Delta\Omega^{MF}[\phi,\eta]\big),
\end{equation}
where 
\begin{equation}
\label{Xi}
\Xi=\int D\eta\int D\phi e^{-\beta\Delta\Omega^{MF}}
\end{equation}
 is the normalization constant and
$\Delta\Omega^{MF}[\phi,\eta]=\Omega^{MF}[\phi,\rho^*]-\Omega^{MF}[0,\rho^*_0]$, where for $\phi=0$ and $\rho^*=\rho^*_0$ the
$\Omega^{MF}[\phi,\rho^*]$ assumes a minimum.
In the  theory studied in Refs.~\cite{ciach:00:0,ciach:05:0,ciach:03:1} $
\Omega^{MF}$ is approximated by
\begin{equation}
\label{Omega}
\Omega^{MF}[\phi,\rho^*] =F_h[\phi,\rho^*] + U[\phi]
-\mu\int_{\bf x} \rho({\bf x}) .
\end{equation}
In the above $\mu$ is the chemical potential of the ions,
$F_h[\phi,\rho^*]=\int_{\bf x}f_h(\phi({\bf x}),\rho^*({\bf x}))$ is the hard-core reference-system Helmholtz
free-energy of the mixture in which the core-diameter $\sigma$ of both
components is the same. We use the Carnahan-Starling (CS) form of
$f_h(\phi,\rho^*)$ in the local-density approximation, as in Ref.\cite{ciach:06:2}.
Finally, the energy in the RPM is given by
\begin{equation}
\label{URPM}
\beta U[\phi]=\frac{\beta^*}{2}
\int_{\bf x}\int_{\bf x'} \frac{\theta(|{\bf x'- x}|-1) }{|{\bf x-x'}|}
 \phi({\bf x}) \phi({\bf x}')
 =\frac{\beta^*}{2}\int_{\bf k}\tilde \phi({\bf k})
 \tilde V (k) 
\tilde \phi(-{\bf k}),
\end{equation}
where $\int_{\bf k}\equiv\int d{\bf k}/(2\pi)^3$, and $\tilde f({\bf k})$ denotes a Fourier transform of $ f({\bf x})$. Contributions to the
electrostatic energy coming from overlapping cores are not included in
(\ref{URPM}).  Here and below the distance is measured in $\sigma$
units, and $\beta^*=1/T^*$ with $T^*$ defined in Eq.(\ref{T*}). The Fourier transform of
$V(x)=\theta(x-1)/x$, where $x=|{\bf x}|$, is
\begin{equation}
\label{V}
\tilde V ( k)=4\pi\cos k /k^2,
\end{equation}
where $k$ is in $\sigma^{-1}$ units.

In the field theory the charge-density fields are considered. The
correlation function for the fields at points  ${\bf x}$ and
${\bf x}'$, called the charge-density correlation function, is given by
\begin{equation}
\label{Charge_cor}
G_{\phi\phi}(|{\bf x}-{\bf x}'|)\equiv
\langle \phi({\bf x}) \phi({\bf x}')
\rangle
=\Xi^{-1}\int D\eta\int D\phi e^{-\beta\Delta\Omega^{MF}}
\phi({\bf x}) \phi({\bf x}').
\end{equation}
The charge-density correlation function defined in
Eq.(\ref{Charge_cor}) is related to the analog of the total charge-charge
correlation function $ h_D(x)$
\cite{stell:95:0,leote:94:0}, and in Fourier representation
this relation is given by
\begin{equation}
\label{h_D}
\frac{\tilde G_{\phi\phi}(k)}{\rho^*}=\tilde h_D(k)\rho^*+1.
\end{equation}
%%
%%%%%%%%%%%%%%%%%%%%%%%%%%%%%%%%%%%%%%%%555
\subsection{Weighted-field approximation}
 Because
$\Delta\Omega^{MF}[0,\eta]$ is strictly local, the field $\eta$ can be
integrated out exactly~\cite{ciach:06:1} and Eq.(\ref{Charge_cor}) takes the
form
\begin{equation}
\label{Charge_cor_eff}
 G_{\phi\phi}(|{\bf x}-{\bf x}'|)
=\Xi^{-1}\int D\phi e^{-\beta{\cal H}_{eff}[\phi]}
\phi({\bf x}) \phi({\bf x}') ,
\end{equation}
 where
\begin{equation}
\label{hef}
\beta{\cal H}_{eff}[\phi]=
\frac{1}{2}\int_{\bf x}\int_{\bf x'}\phi({\bf x})
{\cal C}_{\phi\phi}^0({\bf x}-{\bf x}')\phi({\bf x}')
+\sum_{m=2}^{\infty}\frac{{\cal A}_{2m}}{(2m)!}
\int_{\bf x}\phi^{2m}({\bf x}).
\end{equation}
In the above
\begin{equation}
\label{calC}
{\cal C}_{\phi\phi}^0({\bf x},{\bf x}')=C_{\phi\phi}^0({\bf x},{\bf x}')+
f_2\delta({\bf x}-{\bf x}'),
\end{equation}
where in Fourier representation $C_{\phi\phi}^0$ is given by
\begin{equation}
\label{secder}
\tilde C_{\phi\phi}^0(k)=
\frac{\delta\Delta\Omega^{MF}}
{\delta\tilde\phi({\bf k})\delta\tilde\phi(-{\bf k})}.
\end{equation}
The $ f_2$ and the coupling constants ${\cal A}_{2m}$ are
functions of $\rho_0^*$, whose exact forms depend on  $f_h(\phi,\rho^*)$
\cite{ciach:06:1}, and  have not been determined
yet. Approximate expressions for ${\cal A}_{2m}$ can be obtained in
the weighted field approximation (WF) introduced in
Ref.\cite{ciach:00:0}, and described in more detail in
Refs. \cite{ciach:04:1,ciach:05:0,ciach:06:0}. In the WF the field
$\eta({\bf x})$ is approximated by its most probable form for each
given field $\phi({\bf x})$. Another words, for a given field
$\phi({\bf x})$, the field $\eta({\bf x})$ is determined by the
minimum of $\beta\Delta\Omega^{MF}[\phi,\eta]$, i.e. it assumes the form
\begin{equation}
\eta_{WF}(\phi({\bf x}))=\sum_n\frac{a_n}{n!}\phi({\bf x})^{2n} ,
\end{equation}
where $a_n$ are functions of $\rho_0^*$ \cite{ciach:04:1,ciach:06:2}.
The average density of ions in the WF  is given by
\begin{eqnarray}
\label{14}
\langle \rho^*({\bf x})\rangle=\rho^*_0+
\Xi^{-1}\int D\phi e^{-\beta{\cal H}_{eff}[\phi]}
\eta_{WF}(\phi({\bf x})).
\end{eqnarray}
 The WF approximation is
equivalent to the lowest-order, zero-loop approximation for the exact
${\cal H}_{eff}[\phi]$
\cite{ciach:06:1}. In the WF $ f_2=0$ \cite{ciach:06:1}, and
 ${\cal A}_{2n}$ and $a_n$, given in terms of derivatives of
 $f_h(\phi,\rho^*)$ at $\phi=0$ and $\rho^*=\rho^*_0$, are functions of $\rho^*_0$ \cite{ciach:04:1}.

 In this work we  limit ourselves to
the $\phi^6$ theory, with ${\cal H}_{eff}$ approximated by
\begin{eqnarray}
\label{phi6t}
\beta{\cal H}_{WF}[\phi]=
\frac{1}{2}\int_{\bf x}\int_{\bf x'}\phi({\bf x})
C_{\phi\phi}^0({\bf x}-{\bf x}')\phi({\bf x}')
+\frac{{\cal A}_{4}}{4!}
\int_{\bf x}\phi^{4}({\bf x})+ \frac{{\cal A}_{6}}{6!}
\int_{\bf x}\phi^{6}({\bf x}).
\end{eqnarray}
  The explicit forms of ${\cal A}_4(\rho_0^*)$ and
${\cal A}_6(\rho_0^*)$ in the WF theory with
  the CS approximation for $f_h$ are given in Appendix A~\cite{ciach:06:2}.
In the density interval $0.9<\rho^*_0< 0.1541$ the above
functional has to be supplemented with a term $\propto \phi^8$,
 to ensure stability of $\beta{\cal
  H}_{WF}[\phi]$ for $\phi\to\infty$. In this work  we consider
only $\rho_0^*
  >0.1541$; for this range of densities ${\cal
  A}_{6}(\rho_0^*)>0$.  One should remember that the accuracy of the
  results obtained with the above functional depends on how large are
  the neglected contributions to $\beta{\cal H}_{eff}[\phi]$.

In the
  $\phi^6$ theory the average density in the uniform phase, Eq.(\ref{14}),  
  is approximated by~\cite{ciach:03:1,ciach:06:2}
\begin{equation}
\label{rhoav}
\langle\rho^*({\bf x})\rangle=\rho^*_0+ a_1\langle\phi({\bf x})^{2}\rangle+
\frac{a_2}{2!}\langle\phi({\bf x})^{4}\rangle,
\end{equation}
where the explicit forms of $a_1(\rho_0^*)$ and $a_2(\rho_0^*)$, obtained in Ref.~\cite{ciach:06:2}, are given
 in Appendix A.
On the same level of approximation the number-density correlation
function for ${\bf x}\ne {\bf x}'$ is given by
\begin{eqnarray}
\label{rhorhoav}
G_{\eta\eta}({\bf x},{\bf x}')&\equiv&
\langle \eta_{WF}({\bf x}) \eta_{WF}({\bf x}')\rangle^{con}
=
a_1^2\langle \phi^2({\bf x}) \phi^2({\bf x}')\rangle^{con}
\nonumber \\
&&+
\frac{a_1a_2}{2}\Big(\langle \phi^4({\bf x}) \phi^2({\bf x}')\rangle^{con}
+ ({\bf x}\leftrightarrow {\bf x}')\Big)+
 \frac{a_2^2}{4}\langle \phi^4({\bf x}) \phi^4({\bf x}')\rangle^{con},
\end{eqnarray}
where
\begin{eqnarray}
\label{conn}
\langle f(\phi({\bf x})) g(\phi({\bf x}'))\rangle^{con}\equiv
\langle f(\phi({\bf x})) g(\phi({\bf x}'))\rangle-
\langle f(\phi({\bf x}))\rangle\langle g(\phi({\bf x}'))\rangle.
\end{eqnarray}
%%

%%%%%%%%%%%%%%%%%%%%%%%%%%%%%%%%%%%%%%%%
\subsection{One-loop self-consistent Hartree approximation}
Even for the approximate effective Hamiltonian (\ref{phi6t}),
 $G_{\phi\phi}$ given in Eq.(\ref{Charge_cor_eff}) cannot be calculated
analytically. In the perturbation theory
\cite{amit:84:0,zinn-justin:89:0} $ G_{\phi\phi}$ is given by Feynman
diagrams~\cite{amit:84:0,zinn-justin:89:0} with the $2n$-point vertices
 ${\cal A}_{2n}$. The vertices at
${\bf x}$ and ${\bf x}'$ are connected by lines representing
$G^0_{\phi\phi}({\bf x}-{\bf x}')$, where $G^0$ is inverse to $C^0$ defined in Eq.(\ref{secder}),
and all lines are paired. The
corresponding expressions are integrated over all vertex points, or in
Fourier representation over all $ \tilde G^0_{\phi\phi}(k)$-line
loops, where~\cite{ciach:00:0,ciach:06:2}
\begin{equation}
\label{gau0}
\tilde G_{\phi\phi}^{0-1}(k)=\tilde C_{\phi\phi}^0(k)=
\beta^* \big(S+\tilde V(k)).
\end{equation}
Note that
\begin{equation}
\label{SMF}
S= \frac{4\pi}{x_D^{2}},
\end{equation}
where
\begin{equation}
\label{xD}
x_D=\kappa_D\sigma=\sqrt{4\pi\beta^*\rho^*_0}
\end{equation}
is the dimensionless inverse Debye length.

   The one-loop contribution to $
\tilde C_{\phi\phi}$ (Fig.2a) is proportional to 
${\cal A}_4{\cal G}_0$, where
\begin{equation}
\label{G0}
{\cal G}_0=\int_{\bf k} \tilde
G^0_{\phi\phi}(k).
\end{equation}
%%

%%%%%%%%%%%%%%%%%%%%%
\begin{figure}
\includegraphics[scale=0.5]{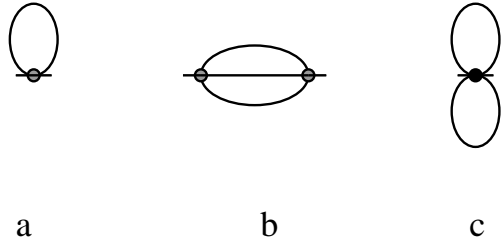}
\caption{Feynman diagrams contributing to $ C_{\phi\phi}$ to
 two-loop order.
The shaded circle and the bullet represent ${\cal A}_4$ and ${\cal A}_6$
respectively. Lines represent $ G^0_{\phi\phi}$. In the
self-consistent theory the lines represent $ G^H_{\phi\phi}$. }
\end{figure}

%%%%%%%%%%%%%%%%%%%%%%%%%%%%%%%%%%%%%%%%%%%%%%%%%%%%%%%%
 Note that Eq.(\ref{gau0})
 can  be written in the form
\begin{equation}
\tilde C_{\phi\phi}^0( k)=\beta^*\big(\tau_0+
\Delta \tilde V(k)\big)
\end{equation}
where
\begin{equation}
\label{tau0}
 \beta^*\tau_0=\frac{1}{\rho_0^*}+\beta^*\tilde V(k_b),
\end{equation}
  $k=k_b$ corresponds to  a minimum of $\tilde V(k)$,
and 
\begin{equation}
\label{delvau}
\Delta \tilde V(k)=\tilde V(k)-\tilde
V(k_b)\simeq_{k\to k_b}v_2(k-k_b)^2+O((k-k_b)^3).
\end{equation}
The integrand in Eq.(\ref{G0}) diverges for $k=k_b$ when
$\beta^*\tau_0=0$.  For $k\to k_b$
Eq.(\ref{G0}) is of a similar form as the corresponding integral in
the Brazovskii theory~\cite{brazovskii:75:0}, and ${\cal G}_0$
diverges for $\beta^*\tau_0\to 0$.   In the effectively
one-loop $\phi^6$ theory (\ref{phi6t}), another contribution to $
\tilde C_{\phi\phi}(k)$ is given by a diagram (Fig.2c) that
is proportional to ${\cal A}_6{\cal G}_0^2$
\cite{ciach:04:1,ciach:05:0}. The symmetry factors of the graphs are
calculated according to standard rules 
\cite{amit:84:0,zinn-justin:89:0}.

 In this work $ \tilde C_{\phi\phi}$ is calculated in the effectively
 one-loop approximation, i.e. as in Ref.\cite{ciach:06:2} the diagrams shown in Fig.2 a,c are
 taken into account.  In the
 self-consistent, effectively one-loop Hartree approximation
the inverse correlation function assumes the form
\cite{brazovskii:75:0,ciach:04:1,ciach:05:0}
\begin{equation}
\label{C_rH}
\tilde C^H_{\phi\phi}(k)=r_0+\beta^*\Delta\tilde V(k)
\end{equation}
where
\begin{eqnarray}
\label{C_r}
r_0\equiv\tilde C^H_{\phi\phi}(k_b)=\beta^*\tau_0+
\frac{{\cal A}_4{\cal G}(r_0)}{2}+
\frac{{\cal A}_6{\cal G}(r_0)^2}{8},
\end{eqnarray}
and
\begin{equation}
\label{calG0}
 {\cal G}(r_0)\equiv\langle \phi({\bf x})^2\rangle=
\int_{\bf k} \tilde G^H_{\phi\phi}(k).
\end{equation}

The integral (\ref{calG0})  diverges because of the integrand
behavior for $k\to\infty$. However, the contribution from $k\to\infty$
is unphysical (overlapping hard cores).  When the fluctuations with
$k\approx k_b$ dominate, which is the case for  $r_0\ll\beta^*v_2k_b^2$, then the main
physical contribution to ${\cal G}(r_0)$ comes from $k\approx k_b$. In
this case
\begin{equation}
\label{aapp}
\tilde C^H_{\phi\phi}(k)\simeq r_0+\beta^*v_2(k-k_b)^2
\end{equation}
and
the regularized integral is \cite{brazovskii:75:0}
\begin{equation}
\label{calG1}
{\cal G}(r_0)=\int_{\bf k} \frac{1}{r_0+\beta^*\Delta\tilde V(k)}
\simeq_{r_0\to 0}\int_{\bf k} \frac{1}{r_0+\beta^*v_2(k-k_b)^2}=
\frac{2a\sqrt T^*}{\sqrt{r_0}},
\end{equation}
where 
\begin{equation}
\label{a}
a=k_b^2/(4\pi\sqrt v_2).
\end{equation}

The self-consistent solution of the pair of equations (\ref{C_r}) and
(\ref{calG1}) can be found analytically. The explicit expression for
$r_0$ is given in Appendix B.

The above results allow for obtaining the number-density correlation
function $G_{\eta\eta}$ at the same level of approximation. From
Eq.(\ref{rhorhoav}) we obtain
\begin{equation}
\label{numcor}
G_{\eta\eta}^H({\bf x},{\bf x}')=\Big(2a_1^2+12a_1a_2{\cal G}(r_0)+
18a_2^2{\cal G}(r_0)^2\Big)
G_{\phi\phi}^{H}({\bf x},{\bf x}')^2+
3!a_2^2G_{\phi\phi}^{H}({\bf x},{\bf x}')^4,
\end{equation}
with the relevant diagrams shown in Fig.3, and the symmetry factors
 calculated according to standard rules~\cite{amit:84:0,zinn-justin:89:0}.
%%%%%%%%%%%%%%%%%%%%%
\begin{figure}
\includegraphics[scale=0.4]{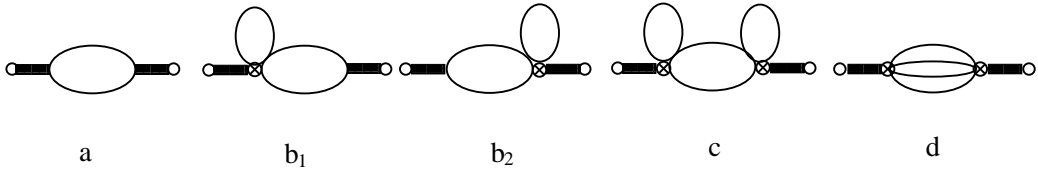}
\caption{Feynman diagrams contributing to
$ G_{\eta\eta}^H({\bf x}_1,{\bf x}_2)$ given in Eq.(\ref{rhorhoav}).
Open circles represent external points ${\bf
x}_1$ and ${\bf x}_2$. Lines  connecting points ${\bf
x}$ and ${\bf x}'$ represent $ G^H_{\phi\phi}({\bf x}-{\bf x}')$. The black box
represents $1/\gamma_{0,2}$, the vertex from which the box and the two
thin lines emanate represents $-\gamma_{2,1}$ and the circle with a
cross inside represents $\Gamma^0_{4,1}$ (see Appendix A, Eqs.(\ref{gamma-mn}) and (\ref{a2ga})). Symmetry factors are
calculated according to standard rules~\cite{amit:84:0,zinn-justin:89:0}. }
\end{figure}
%%%%%%%%%%%%%%%%%%%5
\section{Properties of the correlation functions}
%%%%%%%%%%%%%%%%%%%%%%%%%%%%%%%%%%%%%%%%%%%%%%%%
\subsection{Gaussian approximation}
In Ref.\cite{ciach:03:1} the correlation function $G_{\phi\phi}$ was
studied on the level of the Landau-type theory, where $G_{\phi\phi}$
is approximated by $G^0_{\phi\phi}$.  In Fourier representation the
above approximation for the charge-density correlation function is
given in Eq.(\ref{gau0})\cite{ciach:00:0,ciach:03:1}.  The Fourier
transform $\tilde G_{\phi\phi}(k)$ of the correlation function defined
in Eq.(\ref{Charge_cor_eff}) reduces to $\tilde G_{\phi\phi}^{0}(k)$
given in Eq.(\ref{gau0}) when all $\phi$-dependent terms in
$\beta\Delta\Omega^{MF}$, except from the Gaussian part
$\frac{1}{2}\int_{{\bf k}}\tilde\phi({\bf k})\tilde
C_{\phi\phi}^0(k)\tilde\phi(-{\bf k})$, are neglected.

Let us first focus on the behavior of $\tilde
G^0_{\phi\phi}(k)$  for $k\to 0$.
From Eqs.~(\ref{gau0}) and (\ref{V})
 we obtain
\begin{equation}
\label{still_form}
\frac{\tilde G^0_{\phi\phi}(k)}{\rho^*_0}
\simeq_{k\to 0}\frac{k^2}{k^2+x_D^2},
\end{equation}
where  the dimensionless Debye length $x_D$ is given in Eq.(\ref{xD}).
Note that  Eq.(\ref{still_form}) (with Eq.~(\ref{h_D}))
is consistent with
 the exact second-moment condition of Stillinger and Lovett
 \cite{stillinger:68:0,stell:95:0} and with the Debye-H\"uckel
 result. 

For finite values of $k$ the properties of $\tilde G^0_{\phi\phi}(k)$
agree with the results of other theories~\cite{outhwaite:75:0} only
qualitatively~\cite{ciach:03:1}. The qualitative agreement with the
GMSA results  for $\tilde h_D(k)$ \cite{leote:94:0} is found only in
the phase-space region given by
\begin{equation}
\label{SSMF1}
 S\gg S_{\lambda}\equiv -\tilde V(k_b),
\end{equation}
where $S$ is defined in Eq.(\ref{SSMF}).  The straight line $ S=
S_{\lambda}$ is equivalent to $\tau_0=0$, and represents the boundary
of stability of the functional $ {\cal H}_{WF}$ in the
$(\rho^*_0,T^*)$ phase diagram~\cite{ciach:05:0,ciach:06:2}. At this
line the second functional-derivative of ${\cal H}_{WF}$ vanishes for
the most probable fluctuations~\cite{ciach:01:0,ciach:05:0}. The line
of instability of ${\cal H}_{WF}[\phi]$ is of course the same as the
line of instability of the functional $\Delta\Omega^{MF}$ (see
Eq.(\ref{secder}) and Refs.\cite{ciach:03:1,ciach:05:0,ciach:06:2}).
As shown in Ref.~\cite{ciach:03:1}, the decay length and the amplitude
of $G^0_{\phi\phi}(x)$ both diverge for $ S\to S_{\lambda}$.

 The charge-density correlation function in real-space representation
 is obtained in Ref.\cite{ciach:03:1} by a pole analysis of $\tilde
 G^0_{\phi\phi}(k)$. The Kirkwood line
in the Landau theory~\cite{ciach:03:1} is $S=S_K$, where
%%%%%%%%%%%
\begin{equation}
\label{mon}
S_K=\frac{4\pi\cosh \alpha_{K}}{\alpha_{K}^2},\hskip1cm
\tanh\alpha_{K}=\frac{2}{\alpha_{K}},
\end{equation}
%%%%%%%%%%
and $S_K\approx 11.8$.
In the GMSA and in some other MF theories
 \cite{leote:94:0,outhwaite:75:0} the Kirkwood line is also a straight
 line, but
 in the GMSA its slope was found to be $S_K=4\pi/x_c^2\approx
 8.33$~\cite{leote:94:0}.
For $ S>S_{K}$ there are two inverse decay-lengths, $a_1$ and $a_2$, and
%%%%
\begin{equation}
\label{monodecay}
xG^0_{\phi\phi}(x)=A^{(1)}_{\alpha\beta}e^{-a_1x}+
A^{(2)}_{\alpha\beta}e^{-a_2x}.
\end{equation}
For $S\to S_K^+$ the two imaginary poles of $\tilde
G^0_{\phi\phi}(k)$, $ia_1$ and $ia_2$, merge together and for $ S<
S_{K}$ a pair of conjugate complex poles, $i\alpha_0\pm \alpha_1$,
appears. As a result, $
G^0_{\phi\phi}(x)$ exhibits an  oscillatory behavior for $ S< S_{K}$,
%%%%%%%%%%%%
\begin{equation}
\label{char}
xG^0_{\phi\phi}(x)= -{\cal A}_{\phi\phi}\sin(\alpha_1 x+
\theta)e^{-\alpha_0x},
\end{equation}
%%%%%%%%%%%
with the amplitude ${\cal A}_{\phi\phi}$, the phase $\theta$ and the
inverse lengths $\alpha_0$ and $\alpha_1$ given in
Ref.\cite{ciach:03:1}.  When the $\lambda$-line is approached, both
$\alpha_0$ and ${\cal A}_{\phi\phi}$ diverge as $\propto
(S-S_{\lambda})^{-1/2}$, whereas $\theta\propto (S-S_{\lambda})^{1/2}$
and $\alpha_1\to k_b$.

The number-density correlation function in the above Landau theory for the RPM has
the trivial form
%%%%%%%%%%%%
\begin{equation}
\label{num}
G^0_{\eta\eta}=\gamma_{0,2}^{-1}\delta({\bf x}-{\bf x}'),
\end{equation}
%%%%%%
because the microscopic structure associated with hard spheres in the
coarse-grained description is suppressed.
%%%%%%%%%%%%%%%%%%%%%%%%%%%%%%%%%%%%%%%%%%%%%%%%%%%%%%%%%%%%%%%%%%%%%%%
\subsection{Charge-density correlation function in the self-consistent one-loop
 approximation}
%%%%%%%%%%%%%%%%%%%%%%%%%%%%
In this section we consider the charge-density correlation function
(\ref{Charge_cor}) in the approximation outlined in sec.2.3. As
already explained, our Brazovskii-type approximation relies on two
important assumptions, and therefore it is valid only in a limited
region of the phase diagram. In this section we determine the explicit
form of the correlation functions and also the region of validity of
our results. The approximate expression (\ref{C_rH}) for the inverse
charge-density correlation function can be rewritten in the form
%%%%%%%%%%%%
\begin{equation}
\tilde C_{\phi\phi}(k)=\Big(S^{H}+
\frac{4\pi\cos k}{k^2}\Big)\beta^*.
\end{equation}
%%%%%%%%%%%
 This formula is analogous to Eq.(\ref{gau0}) in
 Ref.\cite{ciach:03:1}, except that $S=T^*/\rho^*_0$ is replaced by
%%%%%%%%%%%%
\begin{equation}
\label{SH}
S^H=T^*r_0-\tilde V(k_b).
\end{equation}
%%%%%%%%%%%

The charge-density correlation function in real space can be obtained
by following the pole analysis of Ref.\cite{ciach:03:1}, just by
substituting $S^H$ for $S$.
The inverse lengths $\alpha_0$ and
$\alpha_1$ are shown in Figs.4 and 5, where the numerical results obtained in
Ref.\cite{ciach:03:1} are used.
%%%%%%%%%%%%%%%%%%%%%%%%%%%%%%%%%%%%%
\begin{figure}
\includegraphics[scale=0.45]{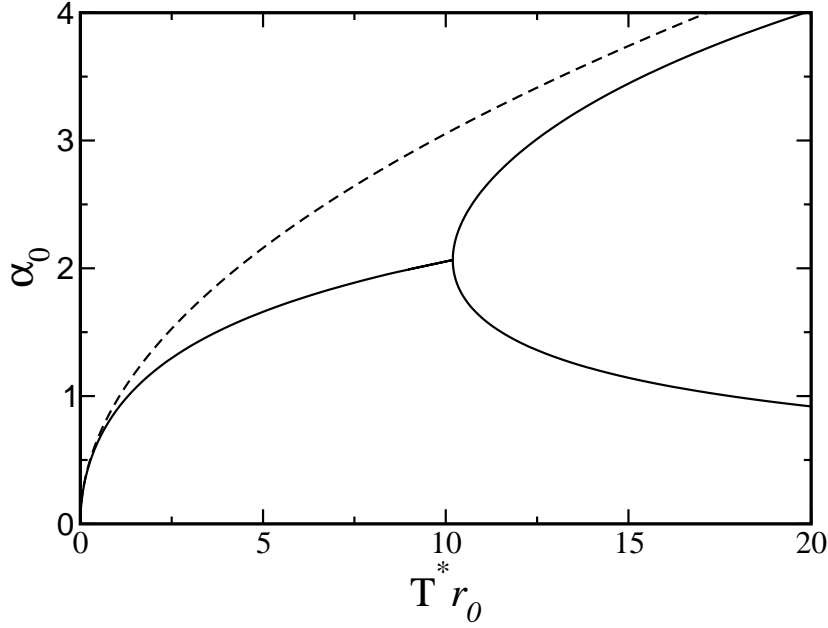}
\caption{The inverse decay length $\alpha_0$ of the
charge-density correlation function, as a function of
$T^*r_0(\rho_0^*,T^*)$, where $r_0(\rho_0^*,T^*)$
is the self-consistent solution of Eqs.(\ref{C_r}) and (\ref{calG1}).
For the explicit form of $r_0$ see Appendix B. The line
$\alpha_0(T^*r_0)$ splits into two branches representing $a_1$ and
$a_2$ at the Kirkwood line $T^*r_0=S_K+\tilde V(k_b)$. Dashed line is
the inverse decay-length obtained from the approximate form of $\tilde
C_{\phi\phi}(k)$ (Eq.(\ref{aapp})).}
\end{figure}
%%%%%%%%%%%%%%%%%%%%%%%%%%%%%%%%%%
   Note that at the $\lambda$-line
 corresponding to the infinite range of correlations ($\alpha_0=0$),
 we find $T^*r_0=0$. From the explicit result for $r_0$ (Appendix B) it follows
 that $r_0>0$ for $T^*>0$. For $T^*>0$ the range of charge-density
 correlations is thus finite, in agreement with the results of the MSA
 and related approximations \cite{stell:95:0,leote:94:0}, as well as
 with simulations.

The dashed line in Fig.4 represents $\alpha_0$ for the approximate form
of $\tilde C_{\phi\phi}(k)$ given in Eq.(\ref{aapp}).  The approximate
form of ${\cal G}$, Eq.(\ref{calG1}), and in turn the form of $r_0$
are based on the above approximation for $\tilde C_{\phi\phi}(k)$.  As
seen form Fig.4, Eq.(\ref{aapp}) yields a good approximation
for $\alpha_0$ when $T^*r_0\le 1$~\cite{ciach:06:2}.
%%%%%%%%%%%%%%%%%%%%%%%%%%%%%%%%%%%%%%%%%%%%%%%%%%%%%%%%%%%%%%%
\begin{figure}
\includegraphics[scale=0.45]{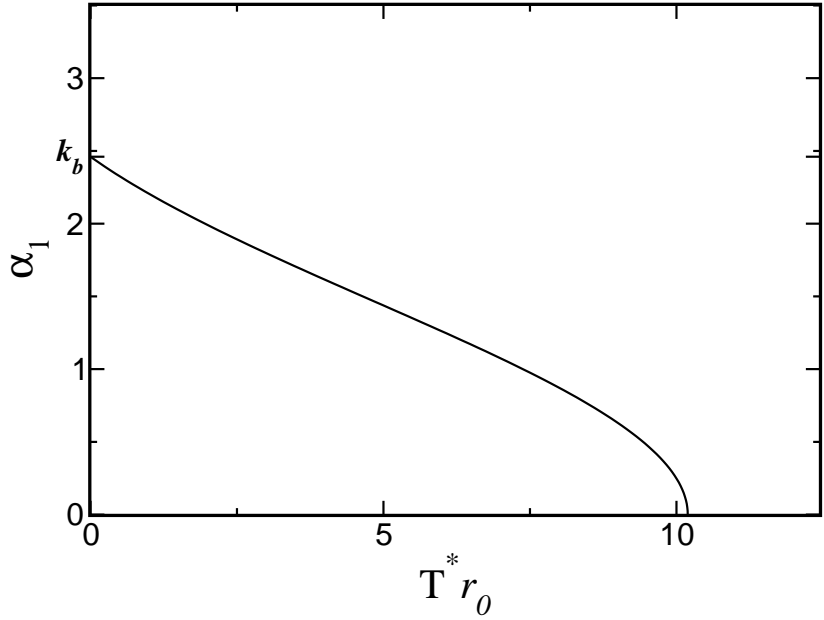}
\caption{$\alpha_1=2\pi/\lambda$, where $\lambda$ is the period
 of damped oscillations of $G_{\phi\phi}(x)$, as a function of
 $T^*r_0(\rho_0^*,T^*)$, where $r_0(\rho_0^*,T^*)$
 is the
 self-consistent solution of  Eqs.(\ref{C_r}) and (\ref{calG1}),
and its explicit form is given in Appendix B. }
\end{figure}
%%%%%%%%%%%%%%%%%%%%%%%%

 Note that the approximation for ${\cal G}(r_0)$ is valid only for
 $r_0\ll\beta^*v_2k_b^2$. For $r_0\gg\beta^*v_2k_b^2$ the fluctuations
 with wavenumbers other than $k\approx k_b$ give a relevant
 contribution to the integral in Eq.(\ref{calG0}), and its value
 depends significantly on the regularization procedure. In addition,
 in the Brazovskii $\phi^4$-theory the two-loop (Fig.2b) and
 higher-order diagrams can be neglected if ${\cal
 A}_4\sqrt{\beta^*v_2}\ll r_0$
\cite{brazovskii:75:0}. From the two above conditions we obtain
the region in the phase diagram,
\begin{equation}
\label{val}
T^*\ll\frac{v_2k_b^4}{{\cal A}_4^2},
\end{equation}
 where Eqs.(\ref{C_r}) and (\ref{calG1}) give accurate results in the
 field theory corresponding to the Hamiltonian (\ref{phi6t}).  The
 line $T^*=v_2k_b^4/{\cal A}_4^2$ is shown in Fig.6 together with the
 coexistence line obtained in Ref.\cite{ciach:06:2}, and the K-line
 (\ref{SHK}) discussed below.  Note that the assumptions of our theory are satisfied near the phase
 coexistence for the high densities $\rho^*_0>0.5$.

 In the approximation derived in sec.3 the  Kirkwood line is given by
%%%%%%%%%%%%
\begin{equation}
\label{SHK}
S^H(\rho^*_0,T^*)=S_K\approx 11.8,
\end{equation}
%%%%%
 where $S^H$ is given in Eq.(\ref{SH}), and the explicit expression
 for
 $r_0(\rho_0^*,T^*)$ is given in Appendix B.  The line
 (\ref{SHK}) is represented by the dotted line in Fig.6. Note that
 this line lies very far from the region of validity of the Brazovskii
 approximation.  Note also that from Eq.(\ref{aapp}) one obtains
 $\alpha_1=k_b$. The approximation (\ref{aapp}) fails of course when
 the Kirkwood line is approached, because at the Kirkwood line and
 beyond $\alpha_1=0$. We conclude that the assumptions of our version
 of the Brazovskii theory are satisfied near the phase coexistence for
 $\rho_0^*>0.5$. When the K-line is approached, the theory becomes less
 accurate, and it fails completely beyond the K-line. On the other
 hand, beyond the K-line the simple Gaussian approximation yields
 results consistent with the exact predictions (see
 (\ref{still_form})).
%%%%%%%%%%%%%%%%%%%%%%%%%%%
\begin{figure}
\includegraphics[scale=1.45]{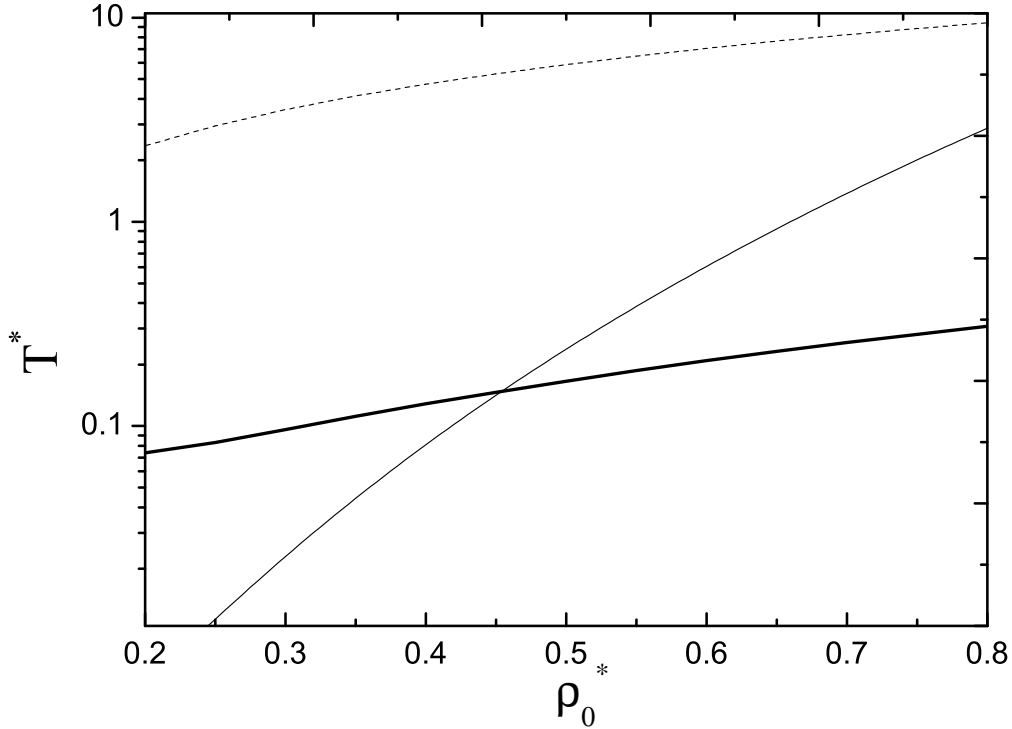}
\caption{Thin solid  line represents $T^*=v_2k_b^4/{\cal A}_4^2$.
 For $T^*\ll v_2k_b^4/{\cal A}_4^2$ the results of the self-consistent
 one-loop approximation with (\ref{calG0}) approximated by
 (\ref{calG1}) are correct. In the rest of the phase diagram either
 the remaining Feynman diagrams (including the one shown in Fig.2b)
 have to be included, or the approximate form of ${\cal G}(r_0)$
 (\ref{calG1}) is not valid.  Thick solid line represents the phase
 transition found in Ref.\cite{ciach:06:2}.  Dotted line is the
 Kirkwood line (\ref{SHK}) in the self-consistent one-loop
 approximation.}
\end{figure}
%%%%%%%%%%%%%%%%%%%%%%%%%%%%%%%%
%%%%%%%%%%%%%%%%%%%%%%%%%%%%%%%%%%%%%%%%%%%%%%%%%%%%%%%%
\subsection{Correlation functions at the coexistence
with the ionic crystal}

The transition between the liquid and ionic crystal was found in
Ref.\cite{ciach:06:2} in an approach consistent with the
self-consistent one-loop approximation outlined in sec.2C.
 Note that in the one-loop approximation $T^*\tilde C_{\phi\phi}(k)$
 depends on the thermodynamic state through a single variable
 $S^H$.
 Since
$\beta^*G^H_{\phi\phi}$ depends on the thermodynamic state only
through $S^H=T^*r_0+S_{\lambda}$ (Eq.(\ref{SH})), we need to know
$T^*r_0$ at the coexistence. From the explicit result for $r_0$
(Appendix B) and the result for the coexistence line $T^*(\rho^*_0)$
obtained in Ref.\cite{ciach:06:2}, we obtain $r_0(\rho^*_0)$ shown in
Fig.7, and in turn we find $T^*r_0\approx 0.2$ for
$0.25<\rho^*_0<0.8$. The inverse lengths along the coexistence are
thus nearly constant and independent of density, and their numerical
values are found to be $\alpha_0\approx 0.97$ and $\alpha_1\approx
2.26$ on the basis of the results obtained in
Ref.\cite{ciach:03:1}. The
approximate results for the amplitude and the phase are
~\cite{ciach:03:1} $-{\cal A}_{\phi\phi}\approx 0.23S^H/\sqrt{S^H-
S_{\lambda}}\approx 0.96$ and $\theta\approx 0.88\sqrt{S^H-
S_{\lambda}}\approx 3.81$ respectively. The correlation functions along
the phase coexistence are shown in Figs.8 and 9.
%%%%%%%%%
\begin{figure}
\includegraphics[scale=0.4]{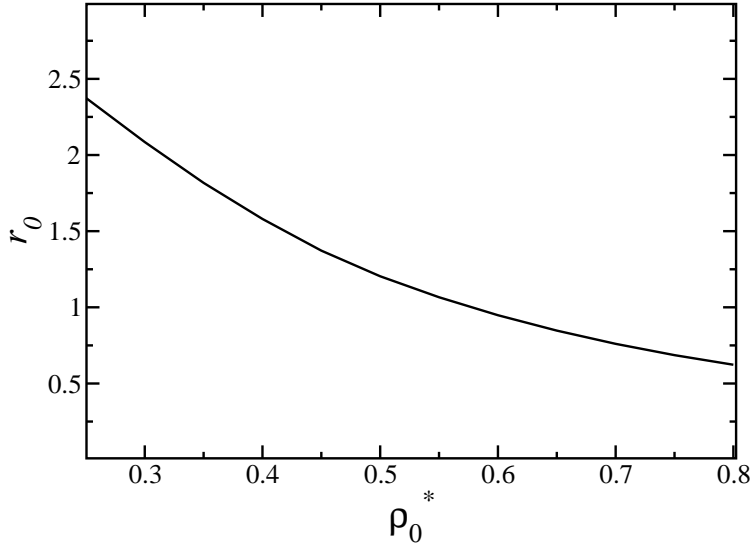}
%\\
\caption{$r_0(\rho^*_0)$ along the coexistence line $T^*(\rho^*_0)$
  between liquid and the charge-ordered phase in the $\phi^6$-theory. 
}
\end{figure}
%%%%

%%%%%%%%%%%%%%%%%%%%%%%%%%%%
\begin{figure}
\includegraphics[scale=1.2]{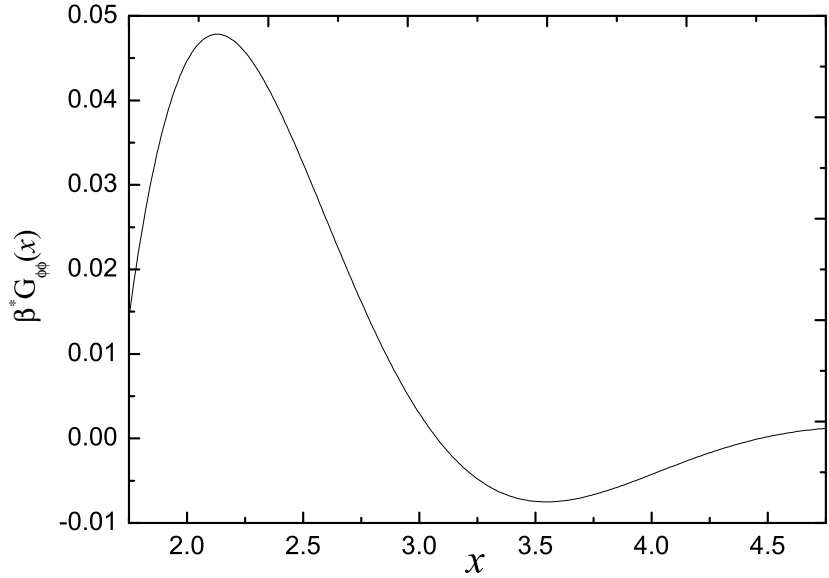}
\caption{Charge-density correlation function $\beta^*G_{\phi\phi}^H(x)$
in the liquid phase for $S^H\approx 1.8$ corresponding to
 the coexistence with the charge-ordered phase.
 $\beta^*G_{\phi\phi}^H(x)$ is
 dimensionless and $x$ is in $\sigma$ units.}
\end{figure}
%%%%%%%%%%%%%
%%%%%%%%%%%%%%%%%%%%%%%%%%%%
\begin{figure}
\includegraphics[scale=1.2]{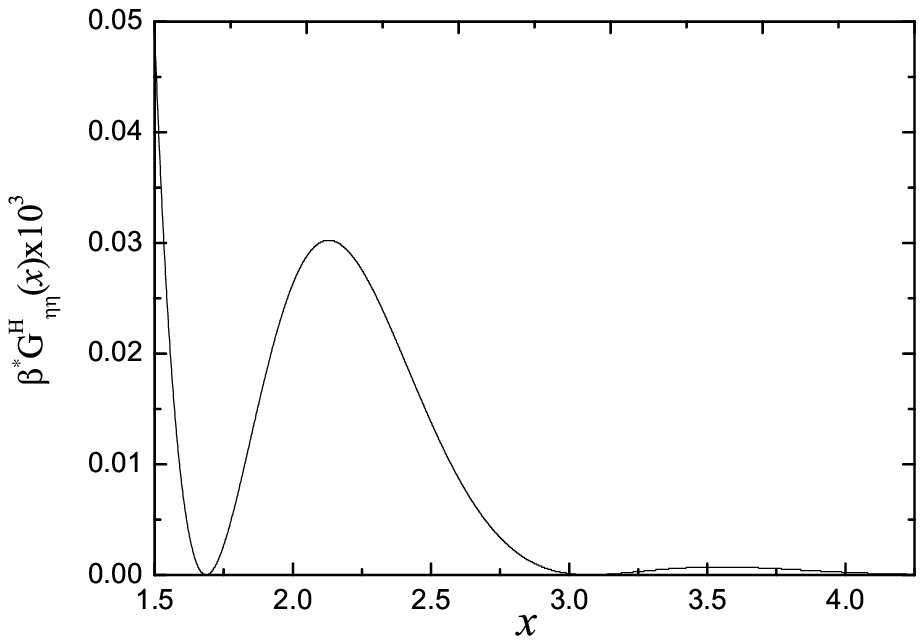}
\caption{Number-density correlation function
$\beta^*G_{\eta\eta}^H(x)$ (Eq.(\ref{numcor})) in the liquid phase at
coexistence with the charge-ordered phase for $\rho^*_0=0.7$. Note
that the number-density correlation function is three orders of
magnitude smaller than the charge-density correlation function.  }
\end{figure}
%%%%%%%%%%%%%%%%%%%%%%%%%%%%%%%%%%%%%%%%

 The inverse lengths $\alpha_0$ and $\alpha_1$ depend on the
 thermodynamic state through a single variable also in the MSA and
 related approximations \cite{leote:94:0}, and
 $x_D=\kappa_D\sigma=\sqrt{4\pi/S}$ is usually chosen. Note that in
 our theory $S^H$ reduces to $S$ in MF; thus, the parameter
 $\sqrt{4\pi/S^H}$ reduces to $x_D$. At the transition
 $\sqrt{4\pi/S^H}\approx 2.25$.  The characteristic inverse lengths in
 the GMSA for $x_D\approx 2.25$ can be read off from Fig.4 of
 Ref.\cite{leote:94:0}, and they are $\alpha_0\approx 1.5$ and
 $\alpha_1\approx 2.1$. More precise values are given in
 Ref.\cite{leote:94:0} for $x_D=4$, where $\alpha_0=1.32$ and
 $\alpha_1=2.89$, and for $x_D=1.5$, where $\alpha_0=2.15$ and
 $\alpha_1=1.22$. Rather good agreement between the two theories is
 obtained, although in our theory the decay length is larger than in
 the GMSA. Similar fair agreement is obtained for the amplitude and
 the phase-shift.

Note that the form of $\beta^*G_{\phi\phi}^H(x)$ is the same along the
phase coexistence, for a large range of density. This result is
consistent with the structure of the ionic crystal coexisting with the
liquid, where the nn distance is also independent of density on the
same level of approximation. This result reflects the fact that the
tendency for ordering in periodic structure follows directly form the
form of the Coulomb potential. 

The number-density correlation function is shown in Fig.9. In our
theory the contribution to the correlation functions associated with
hard-sphere ordering is neglected. The number-density correlation
function shown in Fig.9 results from the coupling between the number-
and the charge-density fluctuations, and is induced by the
charge-ordering, which in turn follows from the form of the interaction
potential. The exact form of the number-density correlation function
should contain the entropic contribution due to the hard-spheres
ordering, in addition to the interaction-induced contribution shown in
Fig.9.
%%%%%%%%%%%%%%%%%%%%%%%%%%%%%%%%%%%%%%%%%%%%%%%%%%%%%%%%%%%%%
\section{Summary}

In this work the fluctuation contribution to the charge- and
number-density correlation functions in the RPM is calculated within
the Brazovskii-type theory. In MF the amplitude and the correlation
length both diverge when the $\lambda$-line is approached. At the
low-temperature side of the $\lambda$-line an instantaneous order
occurs in the majority of the individual microscopic states,
i.a. clusters with oppositely charged nn dominate over randomly distributed
 ions. Averaging over
microscopic states with somewhat different order leads to a
homogeneization of the structure downhomogenizationtion-induced first-order
transition to the ionic crystal. Our purpose here was a verification
if the instantaneous order found on the low-$T$ side of the
$\lambda$-line leaves some traces in the correlation functions. Our
results show that the correlation length and the amplitude of $G_{\phi\phi}(x)$, both
divergent in MF at the $\lambda$-line, show no indication of the MF
singularity when the fluctuation contribution is included. Both
quantities are of order of unity in the whole stability region of the
fluid phase.
%%%%%%%%%%%%%%%%%%%%%%%%%%%%%%%%%%%%%%%%%%%%%%%%%%%%%%%%%%
\begin{acknowledgments}
 This work was supported by the KBN through a research project 1 P02B
 033 26.
\end{acknowledgments}
%%%%%%%%%%%%%%%%%%%%%%%%%%%%%%%%%%%%%%%%%%%%%%%%%%%%%%%%%%%%%%%%%%%%%%%%%%
\section{Appendices}
%%%%%%%%%%%%%%%%%%%%%%%%%%%%%%%%%%%%%%%%%%%%%%%%%%%%%%%%%%%%
\subsection{Coefficients ${\cal A}_4,{\cal A}_6$, $a_{1}$  and $a_{2}$
 in the WF approximation}
\begin{equation}
\label{hypA4}
{\cal A}_4=\gamma_{4,0}-3\frac{(-\gamma_{2,1})^2}{\gamma_{0,2}} ,
\end{equation}
\begin{equation}
\label{hypA6}
{\cal A}_6=\gamma_{6,0}-15\frac{(-\gamma_{2,1})(-\gamma_{4,1})}
{\gamma_{0,2}}
-15\frac{(-\gamma_{2,1})^3(-\gamma_{0,3})}{\gamma_{0,2}^3}-
45\frac{(-\gamma_{2,2})(-\gamma_{2,1})^2}{\gamma_{0,2}^2} ,
\end{equation}
%%
%%%%%%%%%%%%%%%%%%%%%%
where
\begin{eqnarray}
\label{gamma-mn}
\gamma_{2m,n}=\frac{\partial^{2m+n}f_h(\phi,\rho)}
{\partial\phi^{2m}\partial\rho^n}|_{\phi=0,\rho=\rho^*_0}.
\end{eqnarray}
%%%
 In the CS
approximation they assume the explicit forms
\begin{equation}
\label{A4}
{\cal A}_4=-\frac{1-20s+10s^{2}-4s^{3}+
s^{4}}{{\rho_{0}^{*}}^{3}(1+4s+4s^{2}-4s^{3}+s^{4})}
\end{equation}
and
\begin{eqnarray}
\label{A6}
{\cal A}_6&=&\frac{3W(s)}{{\rho^{*}_{0}}^{5}(1+4s+4s^{2}-4s^{3}+s^{4})^{5} },
\end{eqnarray}
where
\begin{eqnarray*}
&&W(s)=3-84s+360s^{2}+2644s^{3}+1701s^{4}-8736s^{5}
\\ \nonumber
&&+11240s^{6}-8304s^{7}+3861s^{8}-1164s^{9}+240s^{10}-36s^{11}+3s^{12}.
\end{eqnarray*}

The coefficients $a_1$ and $a_2$ in Eq.(\ref{rhoav}) are
\begin{equation}
\label{a2}
a_{1}=-\frac{\gamma_{2,1}}{2!\gamma_{0,2}}, \qquad
\frac{a_2}{2}=-\frac{\Gamma^0_{4,1}}{4!\gamma_{0,2}}.
\end{equation}
%%%%
where
\begin{equation}
\label{a2ga}
\Gamma^0_{4,1}=\gamma_{4,1}-\frac{6(-\gamma_{2,2})(-\gamma_{2,1})}
{\gamma_{0,2}}
-\frac{3(-\gamma_{3,0})(-\gamma_{2,1})^2}{\gamma_{0,2}^2}.
\end{equation}
%%%
%%%%%%%%%%%%%%%%%%%%%%%%%%%%%%%%%%%%%%%%%%%%%%%%%%%%%%%%%%%%%%%%%%%%%%%%%%%%%%
\subsection{Explicit expression for $r_0$}
By solving Eqs. (\ref{C_r}) and (\ref{calG1}) we obtain in the case of  $\tau_0<0$ the explicit expression for $r_0$,
\begin{equation}
\label{r0}
r_{0}=\frac{1}{2}\left[ \sqrt{u+v-p/3}+\sqrt{2(h-(u+v)/2-p/3})
+\tau_{0}\right],
\end{equation}

where the following notations are introduced:

\begin{eqnarray*}
u&=&\left(-b_{2}+\sqrt{b_{1}^{3}+b_{2}^{2}}\right)^{1/3}, \qquad
v=\left(-b_{2}-\sqrt{b_{1}^{3}+b_{2}^{2}}\right)^{1/3}, \\
\nonumber b_{1}&=&-\frac{1}{3}\left( \frac{p^{2}}{12}+
8a^{2}T^*{\cal A}_4^{2}\tau_{0}\right),
\quad
b_{2}=\left( -\frac{p}{6}\right)^{3}+\frac{4}{3}pa^{2}T^*{\cal A}_4^{2}\tau_{0}
 -\frac{q}{2}\\
\nonumber
h&=&\left[\left((u+v)/2+p/3\right)^{2}+3\left(u-v\right)^{2}/4
\right] ^{1/2}\\ \nonumber
p&=&-\left(\tau_{0}^{2}+2a^{2}T^*{\cal
A}_6\right),\\
\nonumber
q&=&a^{4}T^{*2}{\cal A}_4^{4}.
\end{eqnarray*}
$\tau_{0}$ and $a$ are defined in Eqs.(\ref{tau0}) and  (\ref{a})
 respectively.
%%%%%%%%%%%%%%%%%%%%%%%%%%%%%%%%%%%%%%%%%%%%%%%%%%

%%%%%%%%%%%%%%%%%%%%%%%%%%%%%%%%%%%%%%%%%%%%%%%%%%
%\bibliographystyle{prsty}
%\bibliography{ions_cit.bib}
%%%%%%%%%%%%%%%%%%%%%%%%%%%%%%%%%%%%%%%%%%%%%%%%%%%%
\end{document}